\documentclass[11pt]{article}
\usepackage{moriond,epsfig}
\usepackage{axodraw}

\def\Journal#1#2#3#4{{#1} {\bf #2}, #3 (#4)}

\def\NPB{{\em Nucl. Phys.} B}
\def\PLB{{\em Phys. Lett.}  B}
\def\PRL{\em Phys. Rev. Lett.}
\def\PRD{{\em Phys. Rev.} D}

\def\be{\begin{equation}}
\def\ee{\end{equation}}
\def\bea{\begin{eqnarray}}
\def\eea{\end{eqnarray}}

\begin{document}
\vspace*{4cm}
\title{$k_T$ FACTORIZATION OF EXCLUSIVE $B$ DECAYS}

\author{HSIANG-NAN LI }

\address{Institute of Physics, Academia Sinica,
Taipei, Taiwan 115, Republic of China}

\maketitle\abstracts{
I compare collinear and $k_T$ factorization theorems
in perturbative QCD, and discuss their application to
exclusive $B$ meson decays.
Especially, I concentrate on the recently measured time-dependent 
CP asymmetry of the $B^0_d\to\pi^+\pi^-$ modes, from which the 
unitarity angle can be extracted.}

\section{Introduction}

Both collinear factorization theorem \cite{BL} and $k_T$ factorization 
theorem \cite{BS,LS} are the fundamental tools of
perturbative QCD (PQCD), where $k_T$ denotes parton transverse momenta.
In these theorems the amplitude for an exclusive process can be calculated 
as an expansion of $\alpha_s(Q)$ and $\Lambda/Q$ with $Q$ being a large 
momentum transfer and $\Lambda$ a small hadronic scale.
Similarly, in the heavy quark limit $m_b\to \infty$, $m_b$ being the $b$ 
quark mass, a $B$ meson decay amplitude can be calculated
as an expansion of $\alpha_s(m_b)$ and $\Lambda/m_b$ \cite{LY1}. A study 
of $B$ meson decays is important for determining Standard Model 
parameters, such as the unitarity angles.

\section{Collinear Factorization vs. $k_T$ Factorization}

We explain how to derive collinear and $k_T$ factorization
theorems for the pion form factor involved in the scattering process
$\pi(P_1)\gamma^*(q)\to\pi(P_2)$. The momenta are chosen in the 
light-cone coordinates as $P_1=(P_1^+,0,{\bf 0}_T)$,
$P_2=(0,P_2^-,{\bf 0}_T)$, and $Q^2=-q^2$.
At leading order, $O(\alpha_s)$, shown
in Fig.~1(a), the hard kernel is proportional to $H^{(0)}(x_1,x_2)\propto 
-1/(x_1P_1-x_2P_2)^2=1/(x_1x_2Q^2)$. Here $x_1$ and $x_2$ are the parton
momentum fractions carried by the lower quarks in the incoming and
outgoing pions, respectively.
At next-to-leading order, $O(\alpha_s^2)$, collinear divergences are 
generated in loop integrals, and need to be factorized into the 
pion wave function. In the collinear region
with the loop momentum $l$ parallel to 
$P_1$, we have an on-shell gluon 
$l^2\sim P_1^2\sim O(\Lambda^2)$ with the hierachy of the components,
$l^+\sim P_1^+ \gg l_T\sim \Lambda 
\gg l^-\sim \Lambda^2/P_1^+$.

\begin{center}
\begin{picture}(100,100)
\Line(10,75)(90,75)
\Line(10,25)(90,25)
\Photon(60,75)(60,100){3}{3}
\Gluon(40,25)(40,75){5}{4}
\Text(10,15)[]{$x_1P_1$}
\Text(90,15)[]{$x_2P_2$}
\Text(50,0)[]{(a)}
\end{picture}\hspace{2cm}
\begin{picture}(100,100)
\Line(9,75)(95,75)
\Line(5,25)(95,25)
\Photon(85,75)(85,100){3}{3}
\Gluon(50,25)(50,75){5}{4}
\Gluon(20,25)(20,75){5}{4}
\Line(35,15)(35,100)
\Text(10,50)[]{$l$}
\Text(20,90)[]{$\phi_\pi^{(1)}$}
\Text(60,90)[]{$H^{(0)}$}
\Text(50,0)[]{(b)}
\end{picture}\par
Fig. 1: (a) Lowest-order diagram for $F_\pi$. (b) Radiative
correction to (a).
\end{center}

An example of next-to-leading-order diagrams is shown in Fig.~1(b).
The factorization of Fig.~1(b) is trivial: one performs the Fierz 
transformation to separate the fermion flows, so that the right-hand
side of the cut corresponds to the lowest-order hard kernel $H^{(0)}$. 
Since the loop momentum $l$ flows into the hard gluon, we have the gluon 
momentum $x_1P_1-x_2P_2+l$ and
\begin{eqnarray}
H^{(0)}\propto\frac{-1}{(x_1P_1-x_2P_2)^2+2x_1P_1^+l^-
-2x_2P_2^-l^++2l^+l^--l_T^2}\;.
\end{eqnarray}
Dropping $l^-$ and $l_T$ as a collinear approximation, the above
expression reduces to
\begin{eqnarray}
H^{(0)}(\xi_1,x_2)\propto\frac{1}{2x_1x_2P_1^+P_2^-+2x_2P_2^-l^+}
\equiv\frac{1}{\xi_1x_2Q^2}\;,
\end{eqnarray}
where $\xi_1=x_1+l^+/P_1^+$ is the parton momentum fraction modified by
the collinear gluon exchange. The left-hand side of the cut then
contributes to the $O(\alpha_s)$ distribution amplitude 
$\phi_\pi^{(1)}(\xi_1)$, which contains the integration over $l^-$ and 
$l_T$. Therefore, factorization to all orders gives a convolution only in
the longitudinal components of parton momentum,
\begin{eqnarray}
F_\pi=\int d\xi_1 d\xi_2\phi_\pi(\xi_1)H(\xi_1,\xi_2)
\phi_\pi(\xi_2)\;.
\end{eqnarray}

In the region with small parton momentum fractions, the hard scale
$x_1x_2Q^2$ is not large. In this case one may
drop only $l^-$, and keep $l_T$ in $H^{(0)}$. This weaker approximation
gives \cite{NL}
\begin{eqnarray}
H^{(0)}(\xi_1,x_2,l_T)\propto
\frac{1}{2(x_1+l^+/P_1^+)x_2P_1^+P_2^-+l_T^2}
\equiv\frac{1}{\xi_1x_2Q^2+l_T^2}\;,
\end{eqnarray}
which acquires a dependence on a transverse
momentum. We factorize the left-hand side of the cut in Fig.~1(b) into
the $O(\alpha_s)$ wave function $\phi_\pi^{(1)}(\xi_1,l_T)$, which
involves the integration over $l^-$. It is understood that the collinear 
gluon exchange not only modifies the momentum fraction, but introduces 
the transverse momentum dependence of the pion wave function.
Extending the above procedure to all orders, we derive the $k_T$
factorization,
\begin{eqnarray}
F_\pi=\int d\xi_1 d\xi_2 d^2k_{1T} d^2k_{2T}
\phi_\pi(\xi_1,k_{1T})H(\xi_1,\xi_2,k_{1T},k_{2T})
\phi_\pi(\xi_2,k_{2T})\;.
\end{eqnarray}

\section{Semileptonic Decays}

Collinear factorization theorem for the semileptonic decay
$B(P_1)\to\pi(P_2)l\nu(q)$ can be constructed in a similar way with the
$B$ meson momentum $P_1=(P_1^+,P_1^-,{\bf 0}_T)$.
Hence, the involved $B\to\pi$ transition form factor is written as 
$F_{B\pi}=\int dx_1dx_2\phi_B(x_1)H(x_1,x_2)\phi_\pi(x_2)$
with the lowest-order hard kernel $H^{(0)}\propto 1/(x_1x_2^2)$. The
parton momentum fractions $x_1$ and $x_2$ are defined via the spectator 
quark momenta $k_1=(x_1P_1^+,0,{\bf 0}_T)$ and $k_2=x_2P_2$ on 
the $B$ meson and pion sides, respectively. Obviously, the above
integral is logarithmically divergent for the asymptotic
model $\phi_\pi\propto x(1-x)$ \cite{SHB}.

There are two options to handle the above end-point singularity:

1. An end-point singularity in collinear factorization implies
that exclusive $B$ meson decays are dominated by soft dynamics. 
Therefore, a heavy-to-light form factor is not calculable \cite{BBNS}, 
and $F_{B\pi}$ should be treated as a soft object, like $\phi_\pi$.

2. An end-point singularity in collinear factorization implies 
its breakdown. $k_T$ factorization theorem then becomes a more
appropriate framework, in which $F_{B\pi}$ is calculable
by meams of the convolution
$F_{B\pi}=\int dx_1 dx_2 d^2k_{1T} d^2k_{2T}
\phi_B(x_1,k_{1T})H(x_1,x_2,k_{1T},k_{2T})
\phi_\pi(x_2,k_{2T})$ \cite{TLS}, with the lowest-order hard kernel
$H^{(0)}\propto
1/\{[x_1x_2m_B^2+({\bf k}_{1T}-{\bf k}_{2T})^2]
[x_2m_B^2+k_{2T}^2]\}$.
It is easy to observe that the above formula does not develop an
end-point singularity.

We emphasize that there is no preference between options 1 and 2 for 
semileptonic $B$ meson decays, since the unknowns $F^{B\pi}$ and 
the meson wave functions are more or less equivalent: from experimental
data one either determines $F^{B\pi}$ in option 1 or $\phi_B$ in option 2.
However, when extending the two options to two-body nonleptonic 
$B$ meson decays, predictions are very different. Options 1 and 2 lead to
the so-called QCD-improved factorization (QCDF) \cite{BBNS} and 
perturbative QCD (PQCD) \cite{KLS,LUY} approaches, respectively.

\begin{center}
\begin{picture}(100,100)
\Line(30,75)(70,75)
\Line(30,35)(70,35)
\CArc(30,55)(20,90,270)
\CArc(70,55)(20,-90,90)
\Line(40,100)(50,75)
\Line(50,75)(60,100)
\Text(50,55)[]{$F_{B\pi}$}
\Text(50,95)[]{$\pi$}
\Text(50,5)[]{(a)}
\end{picture} \hspace{2cm}
\begin{picture}(100,100)
\Line(10,65)(30,50)
\Line(30,50)(10,35)
\Line(30,50)(70,80)
\Line(30,50)(70,20)
\Line(50,50)(90,80)
\Line(50,50)(90,20)
\Gluon(38,56)(50,50){5}{1}
\Text(10,50)[]{$B$}
\Text(30,65)[]{$m_0$}
\Text(50,5)[]{(b)}
\end{picture} \hspace{2cm}
\begin{picture}(100,100)
\Line(30,75)(70,75)
\Line(30,35)(70,35)
\CArc(30,55)(20,90,270)
\CArc(70,55)(20,-90,90)
\Line(40,100)(50,75)
\Line(50,75)(60,100)
\Gluon(60,100)(70,75){5}{2}
\Text(50,55)[]{$F_{B\pi}$}
\Text(50,5)[]{(c)}
\end{picture}\par
Fig. 2: Sources of the strong phase.
\end{center}

\section{Nonleptonic Decays}

Nonleptonic $B$ meson decays occur through a weak effective 
Hamiltonian $H_{\rm eff}$. The decay amplitudes involve several 
topologies: factorizable emission, nonfactorizable emission,
factorizable annihilation, and nonfactorizable annihilation. 
A goal of QCD approaches is to estimate all the topologies of amplitudes.

There are many important quantities involved in two-body
nonleptonic $B$ meson decays. Here we shall discuss only 
the time-dependent CP asymetry of the $B\to\pi\pi$ modes,
\begin{eqnarray}
A_{CP}(t) \equiv {B(\bar{B}^0(t) \to \pi^{+}\pi^{-}) - 
B(B^0(t) \to \pi^{+}\pi^{-}) \over B(\bar{B}^0(t) \to
\pi^{+}\pi^{-}) +B(B^0(t) \to \pi^{+}\pi^{-})}
= S_{\pi\pi}\sin(\Delta m t) - 
C_{\pi\pi}\cos(\Delta m t)\;,
\end{eqnarray}
with $\Delta m$ being the mixing frequency.
The coefficients $S_{\pi\pi}$ and $C_{\pi\pi}$
depend on the weak angle $\phi_2$ and the
tree-over-penguin ratio. We discuss the calculation of the strong 
phase associated with the tree-over-penguin ratio,
and the predictions for $S_{\pi\pi}$ and $C_{\pi\pi}$
in collinear factorization (QCDF) and in $k_T$ factorization
(PQCD).

Sources of the strong phase in QCDF are displayed in Fig.~2. Figure 2(a)
represents the leading $O(\alpha_s^0)$ soft contribution involving the
real $B\to\pi$ form factor. Figure 2(b) is the imaginary annihilation
amplitude from one gluon exchange. Since annihilation occurs through the
scalar penguin operator, it is of $O(\alpha_s r_\chi)$, $r_\chi=2m_0/m_B$
being the chiral enhancing factor with $m_0\sim 1.4$ GeV. Figure 2(c)
denotes the imaginary $O(\alpha_s)$ vertex correction to the four-fermion 
operators. Because of $\alpha_s r_\chi< \alpha_s$ ($r_\chi$ is of order
unity), Fig.~2(c) is the most important source of the strong phase
$\delta$. Compared to the leading contribution of $O(\alpha_s^0)$,
$\delta$ derived from QCDF is expected to be small (and positive). This
is the reason QCDF predicts a smaller and positive $C_{\pi\pi}$ as shown
in the left-hand plot of Fig.~3 \cite{Be02}.

\begin{figure}[t!]
\hspace{-2.0cm}
\epsfig{file=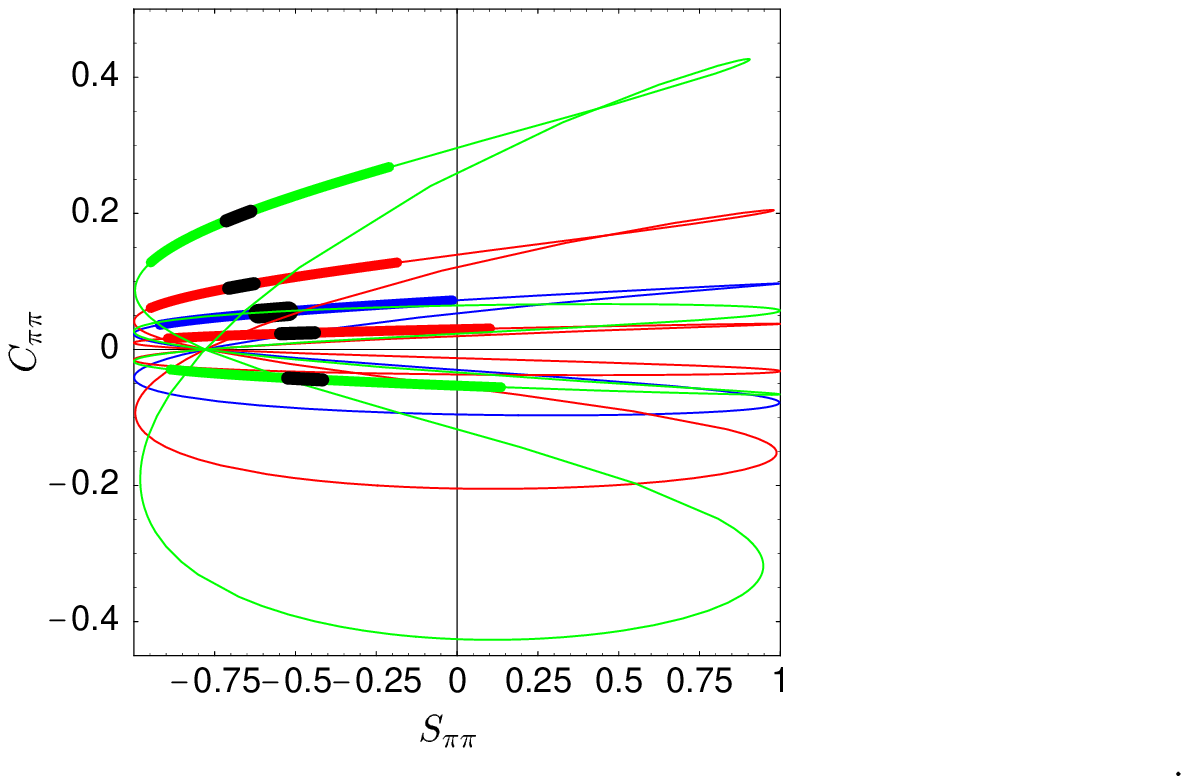,height=2.5in}\vskip -6.5cm\hspace{8.0cm}
\epsfig{file=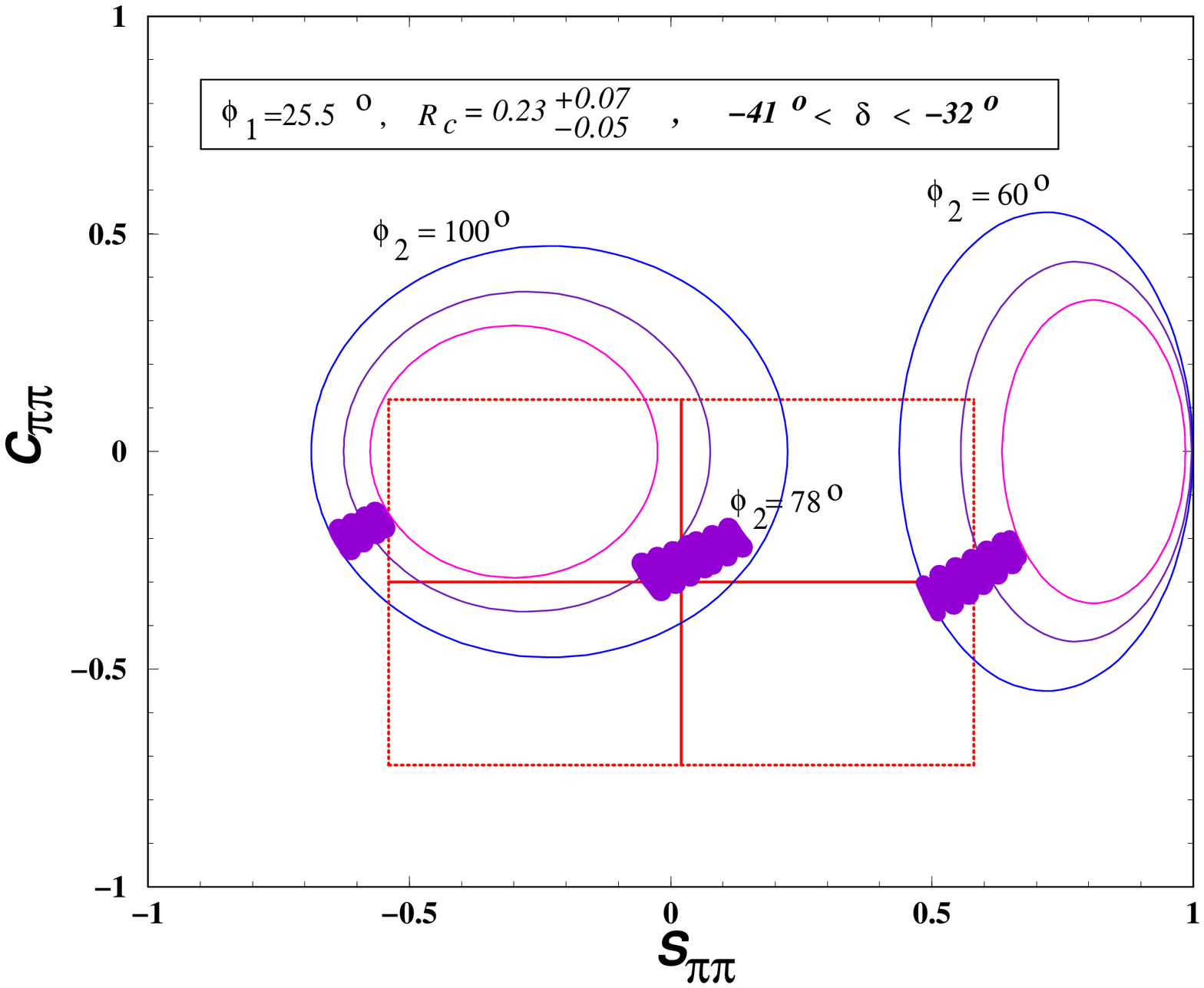,height=2.8in, angle=-90}
\begin{center}
Fig. 3: Correlation between $C_{\pi\pi}$ and $S_{\pi\pi}$
predicted by QCDF and in PQCD.
\end{center}
\end{figure}

In $k_T$ factorization power counting for the topologies of amplitudes
changes. Figure 2(a) represents the $O(\alpha_s)$ calculable $B\to\pi$
form factor, which contains one hard gluon exchange. Figure~2(b) has the
same power counting as in QCDF. Figure~2(c) becomes higher-power, i.e.,
$O(\alpha_s^2)$. Because of $\alpha_s r_\chi\gg \alpha_s^2$, Fig.~2(b)
is the most important source of the strong phase $\delta$. 
It is smaller but comparable to the leading contribution of 
$O(\alpha_s)$. Therefore, $\delta$ derived from PQCD is 
expected to be large (and negative). This is the 
reason PQCD predicts a larger and negative $C_{\pi\pi}\sim -30\%$ as 
shown in the right-hand plot of Fig.~3 \cite{Keum}.
The central value of $S_{\pi\pi}\sim 0$ measured by BaBar
then corresponds to $\phi_2\sim 80^o$. The boundary of the
square represents $1\sigma$ uncertainty, from which 
the range of $\phi_2$, $60^o<\phi_2<100^o$ is extracted \cite{Keum}.

\section{Summary}

In collinear factorization a heavy-to-light transition form factor
exhibits an end-point singularity, while in $k_T$ factorization it is
infrared-finite. Hence, soft dominance is postulated and the
form factor is parametrized as an nonperturbative input in the former.
Hard dominance is postulated and the form factor can be
calculated as a convolution of a hard kernel with meson wave functions in 
the latter. Extending the above theorems to two-body nonleptonic $B$ meson 
decays, QCDF (collinear factorization) prefers a small positive
$C_{\pi\pi}$, direct CP asymmetry in the $B\to\pi\pi$ decays, while
PQCD ($k_T$ factorization) prefers a large negative $C_{\pi\pi}$.
More precise experimental data can soon discreminate the two approaches.

\end{document}